\documentclass[final,3p,times,]{elsarticle}
\usepackage{amssymb}

\newcommand{\vect}[1]{\mbox{\boldmath$#1$}}
\newcommand{\tens}[1]{\mbox{\boldmath$#1$}}
\newcommand{\oper}[1]{\mathbf{#1}}

\newcommand{\dd}{\mathrm d}

\newcommand{\pard}[2]{{\partial_{#2}}{#1}}

\newcommand{\J}{\mathrm J}

\newcommand{\lp}{\left(}
\newcommand{\rp}{\right)}

\newcommand\apj{{Astrophysical Journal }}%
\newcommand\apjs{{Astrophysical Journal Supplements }}%
\newcommand\aap{{Astronomy \& Astrophysics }}%

\journal{Journal of Computational Physics}

\begin{document}

\begin{frontmatter}

\title{LSFEM implementation of MHD numerical solver}

\author[a1,a2]{J. Sk\'ala}
\ead{jskala@physics.ujep.cz}
\author[a2]{M. B\'arta}
\address[a1]{University J. E. Purkinje, \'Ust\'i nad Labem, Czech Republic}
\address[a2]{Astronomical Institute of Czech Academy of Sciences, Ond\v rejov,
  Czech Republic}


\begin{abstract}
Many problems in physics are inherently of multi-scale nature. The issues of MHD
turbulence or magnetic reconnection, namely in the hot and sparse, almost
collision-less astrophysical
plasmas, can stand as clear examples. The Finite Element Method (FEM) with
adaptive gridding appears to be the appropriate
numerical implementation for handling the broad range of scales contained in
such high Lundquist-number MHD problems. In spite the FEM is now routinely used
in engineering practice in solid-state and fluid dynamics, its usage for MHD
simulations has recently only begun and only few implementations exist so far.
In this paper we present our MHD solver based on the Least-Square FEM (LSFEM)
formulation. We describe the transformation of the MHD equations into form
required for finding the LSFEM functional and some
practical issues in implementation of the method. The algorithm was tested on
selected problems of ideal (non-resistive) and resistive MHD.  The tests show
the usability of LSFEM for solving MHD equations.
\end{abstract}


\end{frontmatter}



\section{Introduction}
\label{sect:intro}

Dynamics of magnetized plasma at sufficiently large spatial and temporal scales
can be adequately described by the set of magnetohydrodynamic (MHD)
equations \cite{Priest:1984}.
In many problems we face the situation with high Lundquist (a.k.a. magnetic
Reynolds) number
$$
S\equiv Re_{\rm M}=L\mu_{\rm o} V_{\rm A}/\eta \ ,
$$
where $L$ is the characteristic size of the system, $V_{\rm A}=B/\sqrt{\rho}$
the typical Alfv\'{e}n velocity ($B$ and $\rho$ being the magnetic field
strength and plasma density, respectively) and $\eta$ the electric
resistivity. A direct consequence of the high Lundquist number is a large
separation between the
system size and the dissipation scale. The cascading fragmentation of the current
layer in the magnetic reconnection in solar flares \cite{Barta+:2011a,
Barta+:2011b} can serve as an example of such a multi-scale problem: The span
between the eruption size ($\approx10^5$~km) and the dissipation scale
(1~m -- 10~m) in the practically collision-less coronal plasmas  easily
extends seven orders of magnitude.

In general, there are two approaches how to handle such a broad range of scales.
The first one uses a moderate numerical resolution and models the physics on the
sub-grid (unresolved) scales using some plausible assumptions on
the micro-scale statistical properties (correlations) of the quantities that
define the system (e.g. flow or magnetic field). Among them, e.g., the
Large-Eddy Simulations (LES) \cite{Matais:2001} or Reynolds-Averaged Numerical
Simulations (RANS) \cite{Leschziner:2001} belong to the well known methods
used widely in engineering applications in the fluid dynamics.

The second approach is based on direct
simulations that cover all the scales contained in the problem. Traditionally,
the Adaptive Mesh Refinement (AMR) technique is used with the Finite-Difference/Finite
Volume Methods in order to resolve high-gradient regions locally, keeping the
total number of grid points required for simulation at a manageable level
\cite{Berger+Oliger:1984,Fryxell+:2000,VanDerHolst+Keppens:2007}. Nevertheless,
also this approach has its limitations caused by introduction of artificial
boundaries between fine and coarse meshes. This problem, however, can be cured
by the methods based on unstructured mesh, such as is used in FEM. With this
in mind we have implemented
a FEM-based solver for MHD equations and present it in the
current paper. From various FEM formulations we have chosen the LSFEM because
it is robust, universal (it can solve all kinds of partial differential
equations) and it is efficient -- it always leads to the system of linearized
equations with symmetric, positive definite matrix \cite{Jiang:1998}.
The LSFEM keeps many key properties of the
Rayleigh-Ritz formulation even for systems of equations for which the
equivalent optimization problem (in Rayleigh-Ritz sense) does not exist
\cite{Bochev+Gunzburger:2009}.

Despite of the FEM applications in the fluid dynamics made a substantial development in
the past years, its usage for numerical solution of MHD equations is still
rather rare. For example, the \textit{NIMROD} \cite{Sovinec+:2004} and
\textit{M3D} codes \cite{Jardin+Breslau:2005}
-- based on Galerkin formulation -- belong to a few known implementations
of FEM-based MHD solvers.
Related work also has been done by \citet{Lukin:2008} who implemented the
MHD (and two-fluid) equations within the more general code framework SEL
\cite{Glasser+Tang:2004} based on the Galerkin formulation with high-order Jacobi
polynomials as the basis functions. However, to our knowledge, the LSFEM
implementation of the MHD solver described in the current paper is the first
attempt of this kind.

The paper is organized as follows: First, we briefly describe the underlying
MHD model. Then, the MHD equations are re-formulated in the general
flux/source (conservative) formulation. Temporal discretization, reduction
to the first-order system, and linearization
procedure are described subsequently. Then, the properties of the
least-square formulation of FEM are briefly summarized. Some practical
arrangements of the LSFEM implementation of the MHD solver follows. Finally, the
code is tested on a couple of standardized model problems and the results are
discussed with respect to the intended application of the code to the current-layer
filamentation and decay during the magnetic reconnection in solar eruptions.


\section{MHD equations}
\label{sect:mhd}
The large-scale
dynamics of magnetized plasma can be described by
MHD equations for compressible resistive fluid \cite{Priest:1984}:
\begin{eqnarray}
\label{eq:mhd}
\pard{\rho}{t}+\vect{\nabla\cdot}(\rho\vect u)=0\nonumber\\
\rho\pard{\vect u}{t}+\rho(\vect{u\cdot\nabla})\vect u=-\vect \nabla p+
\vect{j \times B}+\rho\vect{g} \\ \nonumber
\pard{\vect B}{t}=\vect{\nabla\times}(\vect{u\times B})-
\vect{\nabla\times}(\eta\vect j)\\ \nonumber
\pard{U}{t}+\vect{\nabla\cdot S}=\rho\vect{u\cdot g},
\end{eqnarray}
where $\rho$, $\vect{v}$, $\vect{B}$, $U$ are density, macroscopic velocity,
magnetic field, and total energy density, respectively, $\vect{g}$ being
the gravity acceleration.
The energy flux $\vect{S}$ and auxiliary variables
$\vect{j}$ (current density) and $p$
(plasma pressure) are given by the following relations:
\begin{eqnarray}
\nonumber
\nabla\times\vect B=\mu_0\vect j\\
U=\frac{p}{\gamma-1}+\frac{1}{2}\rho u^2+\frac{B^2}{2\mu_0}\\
\nonumber
\vect S=\left(U+p+\frac{B^2}{2\mu_0}\right)\vect u- \frac{(\vect
u\cdot\vect B)}{\mu_0}\vect B+\frac{\eta}{\mu_0}\vect j\times\vect B
\end{eqnarray}

In the (almost) collision-less plasma, in which we are mostly interested, the
classical resistivity usually plays a small role. Instead of that various
microscopical (kinetic) effects influence the plasma dynamics via other terms
in the generalized Ohms law \cite{Buchner+Elkina:2006}. In order to mimic these
processes, whose modeling is beyond the scope of MHD approach, we re-consider
the parameter $\eta$ as a generalized
resistivity, including the effects like wave-particle interactions or off-diagonal
components in the electron pressure tensor into it. As such effects are -- in
general -- observed in the highly filamented, intense current sheets we model
the anomalous generalized resistivity as follows:
\begin{equation}
\eta(\vect r,t)=\left\{
  \begin{array}{lll}
    0 & : & |v_{\rm D}|\le v_{\mathrm{cr}}\\
    C\frac{\left(|v_{\rm D}(\vect r,t)|-v_{\mathrm{cr}}\right)}{v_0}& : &
    |v_{\rm D}|> v_{\mathrm{cr}}
  \end{array}
  \right.
  \label{eq:eta}
\end{equation}
Thus, the non-ideal effects are turned on whenever the current-carrier drift
velocity
\begin{equation}
 v_{\rm D}(\vect r,t)=\frac{|\vect{j}(\vect r,t)|}{e n_\mathrm{e}}
\end{equation}
exceeds the critical threshold $v_{\mathrm{cr}}$.

In order to solve the Eqs.~(\ref{eq:mhd}) numerically, it is convenient to
rescale all the quantities into the dimensionless units. Thus, all the spatial
coordinates are expressed in the characteristic size $L$ and times in Alfv\'en
transit time $\tau_{\rm A}=L/V_{\rm A}$, where $V_{\rm A}=B_0/\sqrt{\rho_0}$ is
a typical Alfv\'en speed. Magnetic field strength $B$ and plasma density $\rho$
are given in units of their characteristic values $B_0$ and $\rho_0$ and similar
scaling holds for the other quantities -- see \cite{Kliem+:2000} or
\cite{Barta+:2011a} for details. From now on we shall use this dimension-less
system.

In order to utilize a more universal LSFEM implementation for more general form
of equations \citep[c.f. with SEL approach][]{Lukin:2008}
the set of MHD Eqs.~(\ref{eq:mhd}) is rewritten into the conservative
(flux/source) formulation:
\begin{equation}
\label{eq:MHDcons}
\pard{\vect{\Psi}}{t}
+\pard{\vect{F}_i(\vect{\Psi},\pard{\vect{\Psi}}{x_j})}{x_i}
=\vect{S}(x_j,\vect{\Psi},t)
\end{equation}
Here the local state vector $\vect{\Psi} = (\rho,\vect{\pi},\vect{B},U)$,
$\vect{\pi}=\rho\vect{v}$ being the momentum density. The flux $\vect{F}$
and the source-term $\vect{S}$ are defined as
\begin{eqnarray}
\label{eq:MHDFluxSource}
  \vect{F}=\lp
  \begin{array}{c}
    \vect{\pi}\\
    \rho\vect{v}\vect{v}-\vect{B}\vect{B}
    +\vect{\hat{I}_{3\times3}}(p+B^2)\\
    \vect{\hat{\epsilon}_{3\times 3} \cdot E}\\
    (h+E_k)\vect{v}+2\vect{E\times B}\
  \end{array}
  \rp \;, \quad
  \vect{S}=\lp
  \begin{array}{c}
     0\\
     \rho\vect{g}\\
     \vect{0}\\
     \vect{\pi \cdot g}
  \end{array}
  \rp \ ,
\end{eqnarray}
where $\vect{\hat{I}_{3\times3}}$ is the $3\times 3$ unit matrix,
$\vect{\hat{\epsilon}_{3\times 3}}$ is the permutation pseudo-tensor,
$\vect{E}=-\vect{v\times B}+\eta\vect{j}$ is the electric field strength. The
the enthalpy and kinetic energy densities are $h=\gamma p/(\gamma-1)$ and $E_k=\rho v^2$,
respectively.


\section{FEM formulation of MHD system}
\label{sect:lsfem}
In general, FEM is formulated for the linear problem
\begin{eqnarray}
  \nonumber
  & &\oper{L}\vect{u}=\vect{f}\qquad {\rm in}\,\Omega \\
  \label{eq:fem} \\
  \nonumber
  & &\oper{B}\vect{u}=\vect{g}\qquad {\rm on}\,\Gamma=\partial\Omega
\end{eqnarray}
where $\mathrm{L}$ is the linear (differential) operator, $\mathrm{B}$ the
boundary operator, $\Omega$ is the domain and $\Gamma$ is the boundary of $\Omega$.

In order to reformulate the system of Eqs.~(\ref{eq:MHDcons}) into the
standardized problem~(\ref{eq:fem}) several steps have to be undertaken. First
of all, we perform the time discretization. We use the standard
$\Theta$-differencing scheme \cite[see, e.g.,][]{Chung:2002}:
\begin{eqnarray}
\label{eq:theta}
\frac{\vect{\Psi}^{n+1}-\vect{\Psi}^{n}}{\Delta t}
=\Theta\left(\vect{S}(x_i,\vect{\Psi}^{n+1},t^{n+1})
-\pard{\vect{F}_i(\vect{\Psi}^{n+1},\pard{\vect{\Psi}^{n+1}}{x_j})}{x_i}\right)
+(1-\Theta)\left(\vect{S}(x_i,\vect{\Psi}^{n},t^{n})
-\pard{\vect{F}_i(\vect{\Psi}^{n},\pard{\vect{\Psi}^{n}}{x_j})}{x_i}\right)
\end{eqnarray}
where parameter $\Theta\in\langle 0, 1\rangle$ controls the implicitness of
the scheme, and $n$ and $n+1$ designate the old and new time-steps,
respectively. The scheme leads to the following semi-implicit equation
\begin{equation}
\label{eq:impl}
\vect{\Psi}^{n+1}+\Theta\Delta t\left(
\pard{\vect{F}_i(\vect{\Psi}^{n+1},\pard{\vect{\Psi}^{n+1}}{x_j})}{x_i}
-\vect{S}(x_j,\vect{\Psi}^{n+1},t^{n+1})\right)=\vect{R}^n \ ,
\end{equation}
where the RHS vector $\vect{R}^n$ consists of components known at old time step.

Since \textit{practical} implementations of LSFEM require first order
system of PDEs \cite{Bochev+Gunzburger:2009, Jiang:1998} we further transform
the system~(\ref{eq:MHDcons}) to the required form introducing a new independent
system variable -- the electric field
\begin{equation}
\label{eq:efield}
\vect{E}=-\vect{v\times B}+\eta\vect{\nabla\times B}
\end{equation}
The procedure is basically analogous to the velocity-vorticity formulation of
the Navier-Stokes equations in the CFD.

A frequent problem in the numerical MHD is a violence of the solenoidal
condition $\rho_\mathrm{m}=\vect{\nabla\cdot B}=0$, where the (dummy) variable
$\rho_\mathrm{m}$ represents the artificial density of the magnetic charge.
The advantage of the LSFEM implementation is that this constraint can be directly
included into the set of the governing equations \cite{Jiang:1998}. Then
assembling the solenoidal condition together
with Eqs.~(\ref{eq:impl}) and (\ref{eq:efield}) we arrive to the
following 1st-order vector equation for our modeled system:
\begin{eqnarray}
\label{eq:mhdExt1}
\left(
\begin{array}{c}
\vect{\Psi}\\
\vect{E}\\
\rho_\mathrm{m}
\end{array}
\right)
+\frac{\partial}{\partial x_i}
\left(
\begin{array}{c}
\tau\vect{F}_i(\vect{\Psi},\vect{E})\\
\eta\vect{G}_i(\vect{\Psi})\\
\vect{H}_i(\vect{\Psi})
\end{array}
\right)
-
\left(
\begin{array}{c}
\tau \vect{S}(\vect{\Psi})\\
\vect{T}(\vect{\Psi})\\
0\\
\end{array}
\right)
=
\left(
\begin{array}{c}
\vect{R^n}\\
\vect{0}\\
0\\
\end{array}
\right) \ ,
\end{eqnarray}
where all the LHS terms are evaluated in the advanced time-step $n+1$.
Here, $\vect{F}(\vect{\Psi},\vect{E})$ and $\vect{S}(\vect{\Psi})$  are given
by Eq.~(\ref{eq:MHDFluxSource}) with $\vect{E}$ considered as an independent
variable now and $\tau\equiv\Theta\Delta t$.
The fluxes $\vect{G}=\vect{-\hat{\epsilon}_{3\times 3} \cdot B}$ and
$\vect{H}=-\vect{B}$ imply from Eq.~(\ref{eq:efield}) and the solenoidal
condition. The source term component
$\vect{T}(\vect{\Psi})=(\vect{\pi}\vect{B}-\vect{B}\vect{\pi})/\rho
-\vect{G(\vect{\Psi})\cdot\nabla\eta}$. We keep the artificial magnetic-charge
density $\rho_\mathrm{m}$ at zero.

Eq.~(\ref{eq:mhdExt1}) can be written in the conservative form similarly as in
Eq.~(\ref{eq:MHDcons})
\begin{equation}
\label{eq:mhdExt2}
\bar{\vect{\Psi}}+\pard{\bar{\vect{F}}_i(\bar{\vect{\Psi}})}{x_i}
-\bar{\vect{S}}(\bar{\vect{\Psi}})=\bar{\vect{R}}
\end{equation}
with the extended state vector, fluxes and source terms in the form
\begin{eqnarray}
\label{eq:mhdExt3}
\bar{\vect{\Psi}}=
\left(
\begin{array}{c}
\vect{\Psi}\\
\vect{E}\\
\rho_\mathrm{m}
\end{array}
\right), \qquad
\bar{\vect{F}}_i=
\left(
\begin{array}{c}
\tau\vect{F}_i(\vect{\Psi},\vect{E})\\
\eta\vect{G}_i(\vect{\Psi})\\
\vect{H}_i(\vect{\Psi})
\end{array}
\right), \qquad
\bar{\vect{S}}=
\left(
\begin{array}{c}
\tau \vect{S}(\vect{\Psi})\\
\vect{T}(\vect{\Psi})\\
0\\
\end{array}
\right), \qquad
\bar{\vect{R}}=
\left(
\begin{array}{c}
\vect{R^n}\\
\vect{0}\\
0\\
\end{array}
\right).
\end{eqnarray}

Since the extended flux $\bar{\vect{F}}$ and source term $\bar{\vect{S}}$
depend
non-linearly on the state vector $\bar{\vect{\Psi}}$, a linearization procedure
has to be applied in order to transform the system~(\ref{eq:mhdExt2})
into the FEM-conforming form~(\ref{eq:fem}). We use the standard Newton-Raphson (NR)
iterations in each time step \cite{Jiang:1998, Sheu+Lin:2004}. Thus, the flux
at the NR iteration $k+1$ can be expressed as \cite{Chung:2002}:
\begin{equation}
\label{eq:NRF}
\bar{\vect{F}}_i(\bar{\vect{\Psi}}^{k+1})=\bar{\vect{F}}_i(\bar{\vect{\Psi}}^{k})
+\left.\pard{\bar{\vect{F}}_i}{\bar{\vect{\Psi}}}\right|_k
\vect{\cdot}(\bar{\vect{\Psi}}^{k+1}-\bar{\vect{\Psi}}^{k})
\end{equation}
and analogous expression holds for the source term. Introducing the Jacobians
\begin{equation}
\label{eq:jacob}
\vect{A}_i\equiv\left.\pard{\bar{\vect{F}}_i}{\bar{\vect{\Psi}}}\right|_k, \qquad
\vect{C}\equiv\left.\pard{\bar{\vect{S}}}{\bar{\vect{\Psi}}}\right|_k
\end{equation}
the final equation for NR iterations reads
\begin{equation}
\label{eq:nrFinal}
\left(\hat{\vect{I}}+\pard{\vect{A}_i}{x_i}
+\vect{A}_i\frac{\partial}{\partial x_i}
+\vect{C}\right)\vect{\cdot}\bar{\vect{\Psi}}^{k+1}=
\bar{\vect{R}}+\left(\pard{\vect{A}_i}{x_i}+\vect{C}\right)
\vect{\cdot}\bar{\vect{\Psi}}^{k}-\bar{\vect{S}}(\bar{\vect{\Psi}}^{k}) \ ,
\end{equation}
where the RHS contains only the terms from the $k-$th iteration of the currently solved
time-step $n+1$ and variables known at the previous step $n$.
Eq.~(\ref{eq:nrFinal}) is already in the form~(\ref{eq:fem}) with
\begin{eqnarray}
\nonumber
\oper{L}=\left(\hat{\vect{I}}+\pard{\vect{A}_i}{x_i}
+\vect{A}_i\frac{\partial}{\partial x_i}
+\vect{C}\right)\\
\label{eq:mhdOper} \\
\nonumber
\vect{f}=\bar{\vect{R}}+\left(\pard{\vect{A}_i}{x_i}+\vect{C}\right)
\vect{\cdot}\bar{\vect{\Psi}}^{k}-\bar{\vect{S}}(\bar{\vect{\Psi}}^{k})
\end{eqnarray}


\section{LSFEM implementation}
\label{sect:impl}
In the least-square formulation of the FEM the problem described by
Eqs.~(\ref{eq:fem}) is transformed to seeks the minimum of the functional
\begin{equation}
\label{eq:femfunc}
 I(\vect{u})=\int_\Omega \lp \oper{L}\vect{u}-\vect{f}\rp^2\dd \Omega+
w\int_\Gamma \lp \oper{B}\vect{u}-\vect{g}\rp^2\dd \Gamma
\end{equation}
where $w$ is appropriate mesh-dependent weighting factor
\cite{Bochev+Gunzburger:2009}. As in other FEM implementations, the solution
is searched for in a limited subspace of functions that are formed as a union of the
piece-wise functions $\vect{u}_e(\vect{x})$ defined on a single, in
our code triangular element, as a
linear combination of the basis functions $\Phi_i(\vect{x})$:
\begin{equation}
\label{eq:basis}
\vect{u}_e(\vect{x})=\vect{u}_e^i \Phi_i(\vect{x})\ ,
\end{equation}
where $\vect{u}_e^i$ can always represent the value of a function
$\vect{u}(x_j^i)$ in a properly selected point (the node) $x_j^i$.
Here $i$ denotes element-wise index of the node. In our code we
use Lagrangian polynomials for basis functions $\Phi_i(\vect{x})$.

Varying the functional (\ref{eq:femfunc}) and inserting the expansion
(\ref{eq:basis}) we arrive to a set of linear algebraic equations for each
internal element in the form
\begin{equation}
 \sum_j \left(\int_{\Omega_e} \oper{L}^T \Phi_i \oper{L} \Phi_j
 \dd\Omega\right)\cdot \vect{u}_e^j
=\int_{\Omega_e} \oper{L}^T\Phi_i \vect{f}\dd\Omega \,,
\label{eq:LSFEM_loc}
\end{equation}
where $\Omega_e\subset\Omega$ is the domain of the $e$-th element in the global
domain $\Omega$. The boundary elements contain additional terms obtained
from the boundary operator (the second term in Eq.~(\ref{eq:femfunc})).
For fast evaluation of local integrals we use Gaussian quadrature
\cite{Chung:2002} in the system of element natural coordinates
\cite{Rathod+:2004}.

Equations~(\ref{eq:LSFEM_loc}) for each element are finally
assembled to a global linear system of equations via mapping the element-wise
node index $j$ to a global node index $N$ described in \cite{Chung:2002}, to obtain
\begin{equation}
\label{eq:FEM_glob}
\sum_N \vect{K}_{iN} \cdot\vect{u}^N = \vect{\hat{f}}^i \ .
\end{equation}
The final matrix $\tens{K}$ is sparse, symmetric and positive definite.
In our code we use preconditioned Jacobi Conjugate Gradient Method (JCGM)
\cite{Press+:2007} for solution of the system~(\ref{eq:FEM_glob}).

The entire algorithm can be summarized as follows:
\begin{itemize}
 \item[--] time loop -- adapt time step size according to CFL condition, check
   final desired time
   \begin{itemize}
    \item[--] linearization loop --
      if $\frac{|\vect{u}^k-\vect{u}^{k+1}|}{|\vect{u}^{k+1}|}<\varepsilon$ or
      maximum iteration count is reached continue to next time step
      \begin{itemize}
       \item[--] assembling stiffness matrix $\tens{K}$ element by element
         \begin{itemize}
          \item[--] integration by Gaussian quadrature
            \begin{enumerate}
              \item compute the operator matrices for each basis function
              \item multiply the operator matrices then add the result
                into stiffness matrix
              \item multiply the operator matrix by the RHS then add result into the
                load vector
            \end{enumerate}
          \item[--] next Gaussian point
         \end{itemize}
       \item[--] next element
      \end{itemize}
    \item[--] find new solution $\vect{u}^{k+1}$ of system~(\ref{eq:FEM_glob})
      by the JCGM
    \item[--] next linearization
   \end{itemize}
 \item[--] next time step
\end{itemize}

Thanks to the iterative nature of the JCGM, the solver algorithm can be rather easily
parallelized via MPI. We decompose the entire domain into subdomains,
splitting the global matrix $\tens{K}$ and the load vector $\vect{\hat{f}}$
into corresponding segments with rather small overlaps related to
internal-boundary nodes shared by both adjacent subdomains. Matrix
multiplications are then
performed only locally (per-process) and, finally, resulting global vectors
are appropriately assembled using MPI operations that transfer the data related
to overlapping nodes only.


\section{Numerical Tests}
\label{sect:numTest}
In order to assess usability and properties of the LSFEM MHD solver we
perform several tests on standardized ideal (non-resistive) and resistive MHD
problems. For all test we use the adiabatic index $\gamma=5/3$, the implicitness
parameter $\Theta=0.5$ (Crank-Nicholson time discretization), and the Courant
number 0.6.


\subsection{Ryu-Jones discontinuity test problem}
\label{subsect:r-j}

\begin{figure}[t]
  \begin{center}
    \includegraphics[width=160mm]{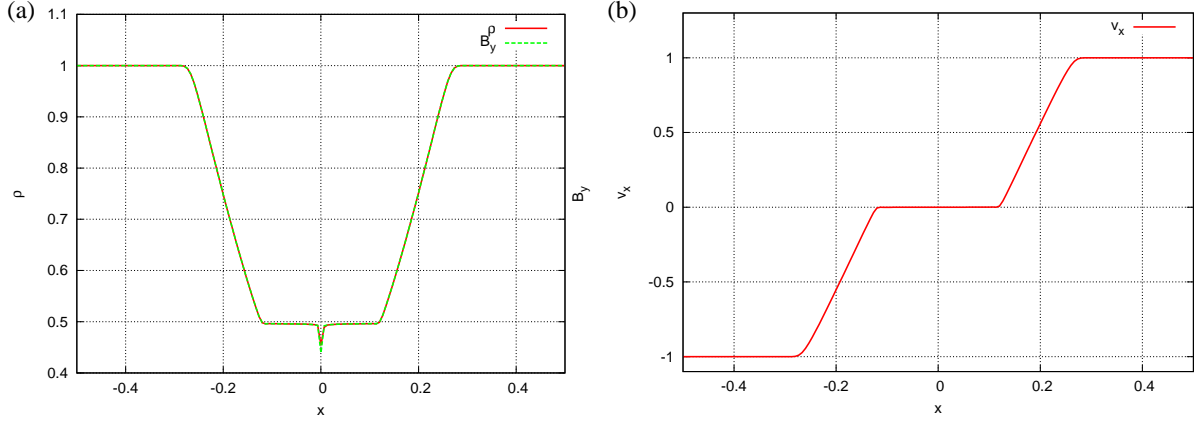}
  \end{center}
  \caption{The LSFEM solution of the MHD shock tube test (Ryu-Jones problem)
    at time $t=0.1$ with the first-order basis functions. (a) density profile
    (red dashed line) and $B_y$ profile along $x$-axis. (b) $v_x$ profile
    along $x$-axis.}
  \label{fig:RJ0LSFEM}
\end{figure}

First, we applied our code onto the standard Ryu-Jones ideal MHD
1D shock/discontinuity problem \cite{Ryu+Jones:1995}.
The initial state is given by prescriptions
$(\rho,v_x,v_y,v_z,B_x,B_y,B_z,E)=(1,-1,0,0,0,1,5,1)$ in the left half, and
$(\rho,v_x,v_y,v_z,B_x,B_y,B_z,E)=(1,1,0,0,0,1,5,0)$ in the right half of the
computational box, respectively. The domain $(-0.5,0.5)$ was divided into 512
elements. We used the first order basis functions to approximate the FEM
solution. The boundary conditions on both ends are of von Neumann type.
Results at time $t=0.1$ are shown in Fig.~\ref{fig:RJ0LSFEM}. They correspond
and could be compared with Fig.~3b in \cite{Ryu+Jones:1995}.

\begin{figure}[t]
 \begin{center}
    \includegraphics[width=160mm]{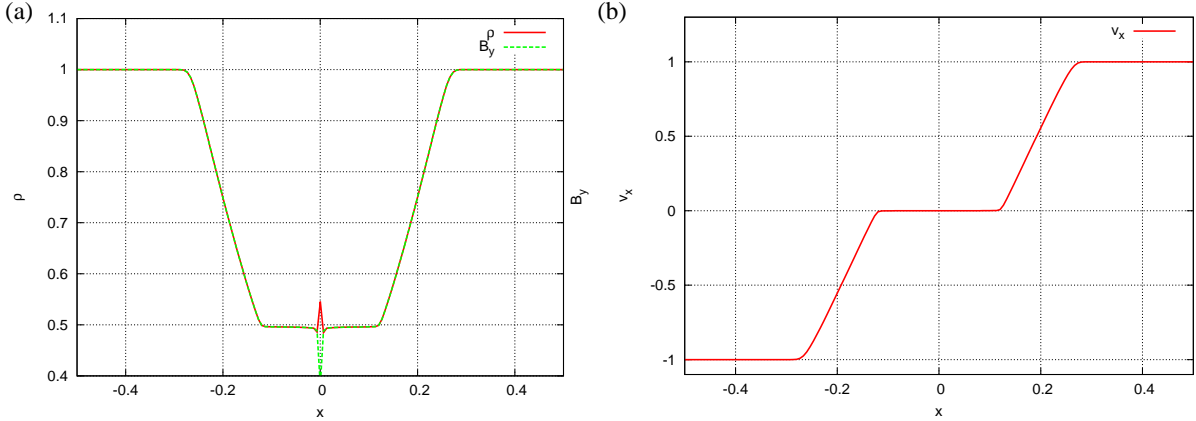}
 \end{center}
  \caption{The LSFEM solution of the MHD shock tube test (Ryu-Jones problem)
   at time $t=0.1$ with the second-order basis functions. Displayed
   quantities are the same as in Fig.~\ref{fig:RJ0LSFEM}.}
 \label{fig:RJ1LSFEM}
\end{figure}

In order to study influence of basis-function order on the approximate
solution we calculate the same test problem, now with the second-order
Lagrange polynomials. All other parameters
are the same as in the previous case displayed in
Fig.~\ref{fig:RJ0LSFEM}. The results are shown in Fig.~\ref{fig:RJ1LSFEM}.


\subsection{Orszag-Tang vortex test problem}
\label{subsect:o-t}
\begin{figure}[t]
 \begin{center}
   \includegraphics[width=160mm]{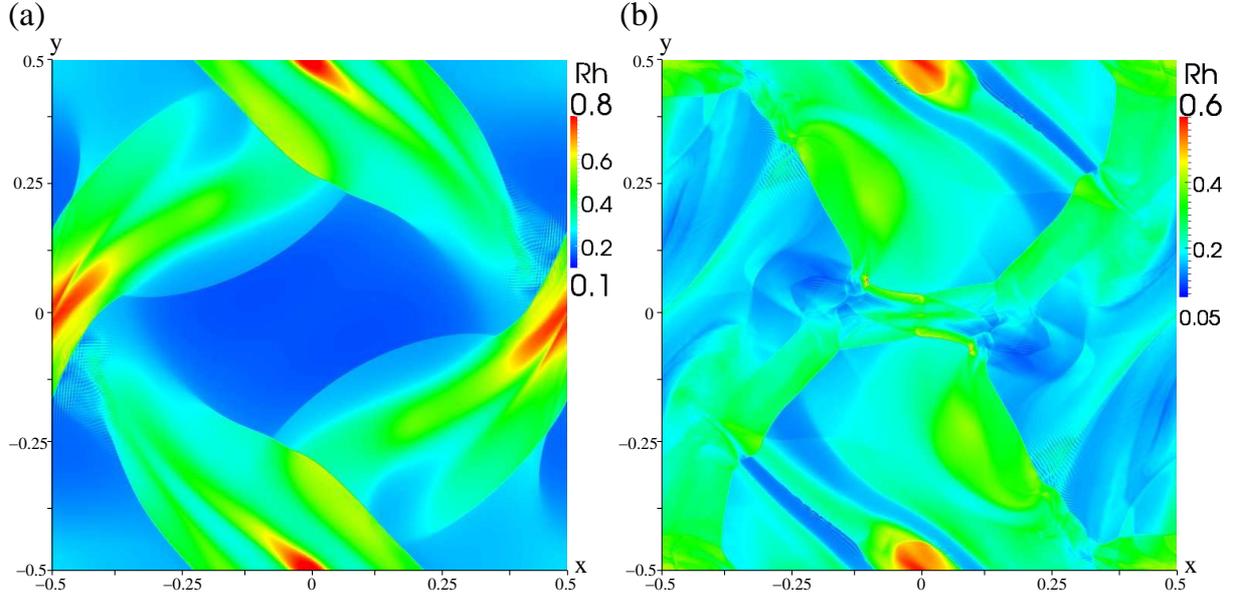}
 \end{center}
 \caption{The Orzsag-Tang vortex. The color coded plasma density is
   displayed at time $t=0.25$ (a) and $t=0.50$ (b).}
 \label{fig:OrszagTangDens}
\end{figure}

\begin{figure}[t]
 \begin{center}
   \includegraphics[width=160mm]{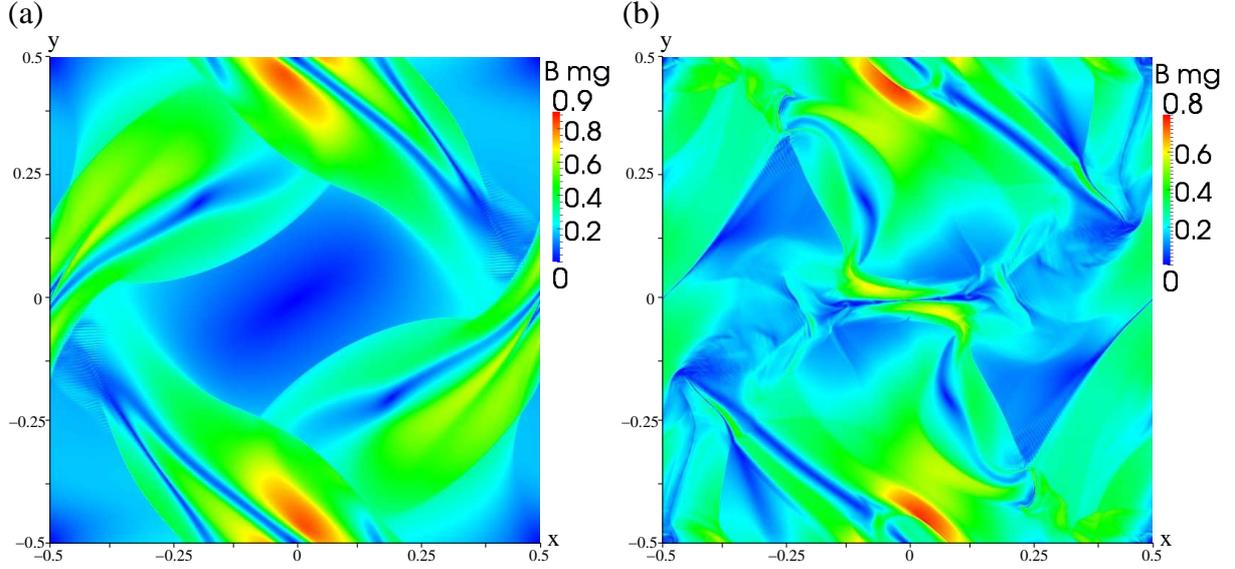}
 \end{center}
 \caption{The Orzsag-Tang vortex. The color coded magnitude of the magnetic field is displayed
   at times $t=0.25$ (a) and $t=0.50$ (b).}
 \label{fig:OrszagTangB}
\end{figure}

A next test we performed standard Orszag-Tang 2D ideal-MHD vortex
problem \cite{Orszag+Tang:1979}. The initial state was given
by the following relations:
\begin{eqnarray}
 \rho &=& p \gamma \nonumber \\
 \pi_x &=& -\sin(2\pi y),\qquad\quad \pi_y = \sin(2\pi x) \nonumber \\
 B_x &=& -\frac{1}{\sqrt{4\pi}}\sin(2\pi y),\quad
 B_y = \frac{1}{\sqrt{4\pi}}\sin(4\pi x)  \nonumber \\
 p &=& \gamma  \frac{1}{4\pi} \nonumber
\end{eqnarray}
The computational domain $1.0\times 1.0$ was discretized by $2\times
640\times 640$ triangular elements. We apply periodic boundary conditions at all
boundaries. The first-order basis functions were used in this simulation.
Results in Figs.~\ref{fig:OrszagTangDens} and~\ref{fig:OrszagTangB}
show the plasma density and the magnitude of the magnetic field, respectively, at times
$t=0.25$ (a), and $t=0.50$ (b).

\subsection{Resistive decay of a cylindric current}
\label{subsect:j-decay}
In order to assess the applicability of our code to the solutions of non-ideal
(resistive) MHD problems and to estimate its numerical resistivity we performed
a following test: At the initial state $t=0$ a cylindrical current
$\vect{j}(r)=(0,0,j_z(r))$ with
\begin{equation}
\nonumber
j_z(r)=\left\{
  \begin{array}{lll}
    j_0 \J_0(x_N \frac{r}{r_0}) & : & r \le r_\mathrm{0}\\
    0 & : & r> r_\mathrm{0}
  \end{array}
  \right.
\end{equation}
flows through a plasma of a uniform density $\rho_0$. Here, $j_0=1$ is the
amplitude of the current density on the cylinder axis, $r_0=1$ is the cylinder
radius, and $x_N\approx 2.40$ is the first null of the Bessel function
of the 0th order $\J_0(x)$. The resistivity inside the
cylinder ($r \le r_\mathrm{0}$) is uniform $\eta=\eta_0=0.1$, outside $\eta=0$. In order to
be able to compare the numerical results with an analytical solution and to split
advective and resistive properties of the code we set all velocities to zero
at $t=0$ and the density to a very high value $\rho_0=10^7$ to keep the plasma
in rest. In the limit $\rho \rightarrow \infty$ the MHD system~(\ref{eq:mhd})
effectively reduces into the diffusion equation
\begin{equation}
\nonumber
\pard{\vect B}{t}+\vect{\nabla\times}(\eta\vect j)=\vect{0}
\end{equation}
whose analytical solution for our initial state keeps the form
$\vect{j}(r,t)=(0,0,j_z(r,t))$ with
\begin{equation}
\label{eq:jt_anal}
j_z(r,t)=j_0 \J_0(x_N \frac{r}{r_0}) \exp(-\gamma t)+
\frac{j_0}{2\pi x_N} \J_1(x_N)\left[ 1-\exp(-\gamma t)\right]\delta(r-r_o)\ ,
\end{equation}
where $\J_1(x)$ is the Bessel function of the 1st order and $\delta(x)$ is the
Dirac delta function. The decrement $\gamma$ reads
\begin{equation}
\label{eq:decrement}
\gamma=\eta\left(\frac{x_N}{r_0}\right)^2\ .
\end{equation}
The second term in Eq.~(\ref{eq:jt_anal}) represents an induced surface current
that compensates resistive decrease of the current density inside the column to
keep the magnetic field in the outer super-conducting domain constant. The
corresponding magnetic field is of the form $\vect{B}=(0,B_{\phi},0)$ where
\begin{equation}
\nonumber
B_{\phi}(r,t)=j_0\frac{r_0}{x_N} \J_1(x_N \frac{r}{r_0}) \exp(-\gamma t)
\end{equation}
for internal ($r\le r_0$) region and
\begin{equation}
\nonumber
B_{\phi}(r,t)=j_0\frac{r_0}{x_N} \J_1(x_N)
\end{equation}
for the outer space.

\begin{figure}[t]
 \begin{center}
    \includegraphics[width=160mm]{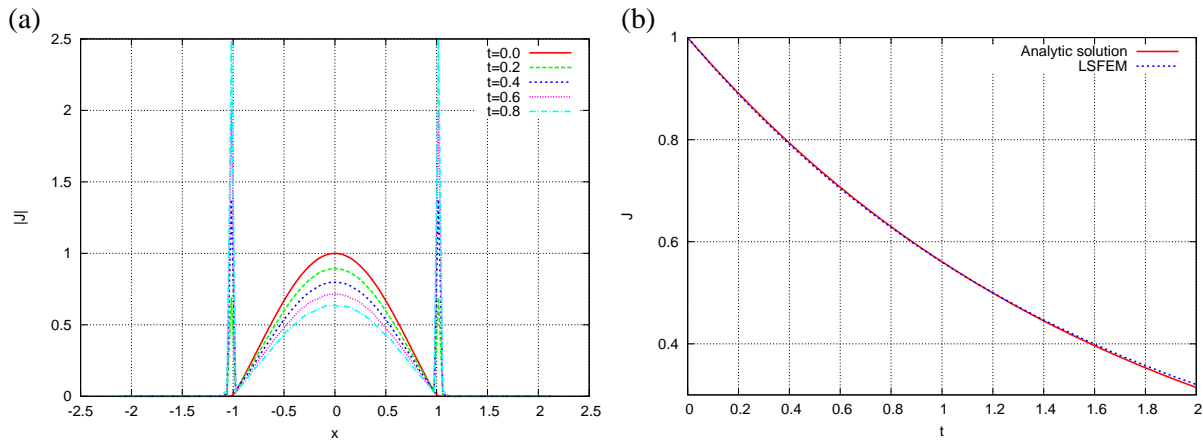}
 \end{center}
 \caption{A resistive decay of a cylindrical current density with time. (a)
   profiles of $j_z(x,0,t)$ at five subsequent times. (b) The time profile of
   $j_z(0,0,t)$ - comparison of numerical and analytical
   [Eq.~(\ref{eq:jt_anal})] solutions.}
 \label{fig:resistive}
\end{figure}

Computational domain is divided into a homogeneous mesh of $2\times 512\times 512$
triangles in our numerical test. We use the first order basis functions to
approximate the numerical solution. Free boundary conditions were applied
on all boundaries. The results of this test are shown in Fig.~\ref{fig:resistive}.
Fig.~\ref{fig:resistive}(a) shows time evolution of the current density
profile along $y=0$ for five subsequent time instants. Resistive
decrease of $j_z$ inside the column accompanied by formation of the induced
surface current
are well visible. Fig.~\ref{fig:resistive}(b) shows a comparison of
numerical and analytical solutions for time evolution of the current density
$j_z(x,y,t)$ at $x=0, y=0$.


\section{Discussion and conclusions}
\label{sect:conclusion}

The FEM represent an alternative to FDM/FVM
that are traditionally used for solution of MHD problems in astrophysics. Its
attractivity implies from its unstructured mesh that allows for appropriate local
refinement without formation of qualitative internal boundaries between the fine
and coarse meshes. This property makes it very useful for handling the
multi-scale problems, for example the problem of magnetic reconnection in solar
flares \cite{Barta+:2011a} (and other large-scale systems) or MHD turbulence.

With this intention in mind we have developed the LSFEM implementation of a MHD
solver whose descriptions and preliminary results from its application to the
standardized test problems are presented in this paper.

To sum up the main points of our implementation: Transformation of the MHD
equations~(\ref{eq:mhd}) to the standard FEM problem~(\ref{eq:fem}) involves
several steps: (i) Standard $\Theta$- time discretization, (ii) Decrease of the
order of the system of equations by introduction of a new variable -- electric
field strength, and (iii) Newton-Raphson linearization. The possibility to
include the solenoidal condition $\vect{\nabla\cdot B}=0$ directly into the system
of equations certainly belongs to advantages of LSFEM formulation, as well as a
natural involvement of the boundary conditions. The element-by-element assembling of
the global stiffness matrix and the iterative nature of JCGM solver
allow for rather easy and efficient MPI parallelization. Integrals over
elements are efficiently performed via Gauss quadrature.

We performed several standardized tests focused on an ideal and resistive MHD.
The LSFEM MHD solver quite closely reproduces results published for the Ryu-Jones shock
tube problem \cite{Ryu+Jones:1995}. Small spurious oscillations appear around
the points where the first derivative of an analytical solution does not
exist. Choice of the higher-order basis functions makes the situation even
slightly worse.

Similar feature can be seen in the results from the Orszag-Tang vortex test problem.
While the large-scale dynamics agree well with those obtained from the 'gauge'
codes, small oscillations accompanying the shocks are visible again. These
effects are caused by the least squares curve fitting approach
\cite{Bochev+Gunzburger:2009}. We believe that it can be cured by an introduction
of a small background resistivity and local refinement of the mesh around the
discontinuities, with the element size corresponding to the resistivity-controlled
(magnetic) Reynolds number. Such approach is fully in line with the intended usage
of the code for detailed studies of the current sheet filamentation and
fragmentation in a large-scale magnetic reconnection in solar flares. Indeed,
in the solar corona we have a very small background classical resistivity due to (rare)
collisions between electrons and ions as well. Hence, having mesh around the
filamenting current sheet locally refined as much as possible we can set the
background (physical) resistivity accordingly and approach thus the realistic
Lundquist number in the solar corona.

Finally -- with the intended usage of the code in mind -- we have tested the
properties of our implementation for solution of the resistive problems. In order to
get a comparison with an analytical solution we have 'frozen' the plasma dynamics by
setting high matter density and we concentrated on a purely diffusive problem. The
results show a rather good agreement with the analytical solution. Namely, the
induced surface current density is located only at a few elements and did not
diffuse further with time. This is an important result for intended future studies
of the current-sheet filamentation in the flare reconnection.

The tests show basic applicability of our LSFEM implementation of the MHD solver
for a solution of selected problems. At the same moment they reveal the necessity
to involve both the adaptive spatial refinement (it has already been implemented)
and adaptive change of the order of basis functions over selected elements
(h-p refinement). These features will be implemented into our code in a near future.


\section*{Acknowledgments}
This research was supported by the grants P209/12/0103 (GA \v{C}R), P209/10/1680
(GA \v{C}R), and the research project RVO:67985815. M.B. acknowledges support of
the European Commission via the PCIG-GA-2011-304265 project financed in
frame of the FP7-PEOPLE-2011-CIG programme.




\end{document}